\documentclass[12pt,a4paper]{article}
\usepackage[T1]{fontenc}
\usepackage{amsmath}
\usepackage[bitstream-charter]{mathdesign}
\usepackage{graphicx,breqn}
\usepackage{bbm,float,longtable,mathtools,subcaption}
\usepackage{bm,algorithm,hyperref,multirow,xr}
\usepackage[toc,page]{appendix}
\usepackage{theoremref,enumitem,authblk}
\usepackage[noend,compatible]{algpseudocode}
\usepackage{apacite,color,soul,xcolor,ntheorem,lipsum,lscape}
\usepackage[round]{natbib}

\usepackage[left=2.5cm,right=2.5cm,top=2.5cm,bottom=2.5cm]{geometry}

\theorembodyfont{\itshape}

\newtheorem*{Proof*}{Proof}

\newtheorem{lemma}{Lemma}

\allowdisplaybreaks

\makeatletter
\newenvironment{breakablealgorithm}
  {
   \begin{center}
     \refstepcounter{algorithm}
     \hrule height.8pt depth0pt \kern2pt
     \renewcommand{\caption}[2][\relax]{
       {\raggedright\textbf{\ALG@name~\thealgorithm} ##2\par}%
       \ifx\relax##1\relax 
         \addcontentsline{loa}{algorithm}{\protect\numberline{\thealgorithm}##2}%
       \else 
         \addcontentsline{loa}{algorithm}{\protect\numberline{\thealgorithm}##1}%
       \fi
       \kern2pt\hrule\kern2pt
     }
  }{
     \kern2pt\hrule\relax
   \end{center}
  }
\makeatother

\numberwithin{equation}{section}

\title{ Estimation of Constrained Mean-Covariance of  Normal Distributions}

\author[1]{Anupam Kundu}
\author[1,2]{Mohsen Pourahmadi}
\affil[1]{\it Department of Statistics, Texas A\&M University}

\date{}
\makeatother
\begin{document}
\maketitle
\footnote{Author address: Department of Statistics, Texas A\&M University, 3143 TAMU, College Station, TX 77843-3143\\Email: $^1$akundu@stat.tamu.edu (Corresponding Author), $^2$pourahm@stat.tamu.edu}

\noindent \textbf{Abstract:} Estimation of the mean vector and covariance matrix is of central importance in the analysis of multivariate data. In the framework of generalized linear models, usually the variances are certain functions of the means with the normal distribution being an exception. We study some implications of functional relationships between covariance and the mean by focusing on the maximum likelihood and Bayesian estimation of the mean-covariance under the joint constraint $\bm{\Sigma}\bm{\mu} = \bm{\mu}$ for a multivariate normal distribution. A novel structured covariance is proposed through reparameterization of the spectral decomposition of $\bm{\Sigma}$ involving its eigenvalues and $\bm{\mu}$. This is designed to  address the challenging issue of positive-definiteness and to reduce the number of covariance parameters from quadratic to linear function of the dimension.  We propose a fast (noniterative) method for approximating  the maximum likelihood estimator by maximizing a lower bound for the profile likelihood function, which is concave. We use normal and inverse gamma priors on the mean and eigenvalues, and approximate the  maximum aposteriori estimators by both MH within Gibbs sampling and a faster iterative method. A simulation study shows good performance of our estimators.\par

\vspace*{1cm}

\noindent\textbf{Abstract Keywords:} Joint Mean-Covariance Estimation, Structured Covariance, Shrinkage Inverse Wishart, Normal-Inverse Gamma, Metropolis-Hastings within Gibbs Sampling.

\section{Introduction}

Mean and covariance estimation of multivariate normal distribution are essential in almost every area of classical multivariate statistics \citep{bibby1979multivariate}. The range of modern applications includes astrophysics \citep{hamimeche2009properties}, economics \citep{ledoit2004honey}, environmental sciences \citep{eguchi2010priori}, climatology \citep{guillot2015statistical} and genetics \citep{schafer2005shrinkage}. \par

Estimation under joint constraints on the mean vector and covariance matrix of data from a  $N_p(\bm{\mu},\bm{\Sigma})$ distribution is relatively uncommon in
 multivariate statistics  \citep{bibby1979multivariate}. It is well-known that estimation of a covariance matrix alone  is a daunting task because of some standard constraints e.g. 1) positive definiteness and 2) the number of unknown covariance parameters growing quadratically with the dimension. Many strategies are developed to bypass the notorious positive definiteness e.g. spectral, Cholesky decomposition and factor models see \cite{chiu1996matrix, pourahmadi1999joint, fan2008high}. Here due to the nature of the constraint on the mean and covariance, we rely on the spectral decomposition and introduce  a new class of structured covariance matrices with several desirable properties. To the best of our knowledge this class of covariance matrices seems to be new and has not been studied before.
 
   We consider  the following joint constraint on the mean vector and covariance matrix of a multivariate normal distribution:
\begin{align}
\label{ESAGCon}
\bm{\Sigma}\bm{\mu} &= \bm{\mu},  \qquad |\bm{\Sigma}| = 1,
\end{align}
which appeared first in \citep{paine2018elliptically} in the context of spherical data. 
Interestingly, the first constraint  forces the mean vector to be an eigenvector of the covariance matrix corresponding to the eigenvalue one and is more consequential. The second which constraints the product of the remaining eigenvalues is less stringent and can be realized by rescaling, so that without loss of generality it is ignored from here on.

Nevertheless, the joint constraint on the mean and covariance matrix will definitely impact their estimators and the shape of the contours of the multivariate normal density function as gleaned from  the spectral decomposition of the covariance matrix:
 \begin{align}
 \label{SDSigma}
 \bm{ \Sigma}=\bm{PDP}^{\top}&=\sum_{i=1}^p \lambda_i \bm P_i \bm P_i^{\top}=\sum_{i=1}^{p-1} \lambda_i \bm P_i \bm P_i^{\top}+\bm{\mu}\bm{\mu}^{\top}
 \end{align} where $\bm D=\text{diag}(1,\lambda_1,\lambda_2,\dots,\lambda_{p-1})$ is the diagonal matrix of ordered  eigenvalues
   other than 1 and $\bm P=[\bm{\mu}, \bm P_1,\bm P_2,\dots,\bm P_{p-1}]$ is an orthogonal matrix of eigenvectors.

   Though these constraints arise in the
context of directional data analysis \citep{paine2018elliptically}, they seem to resonate with some
deep classical issues in statistical estimation theory. Note that the presence of the quadratic term $\bm{\mu}\bm{\mu}^{\top}$ in (\ref{SDSigma}) has resemblance with classical  estimation in the  $N(\theta,\theta^2)$ distribution.  It also encourages modeling the covariance matrix as a parsimonious quadratic function of the mean vector similar to \cite{hoff2012covariance}.

 We propose  a simple and novel  structured covariance model based on the spectral decomposition (\ref{SigStruc}) which overcomes several challenges in covariance estimation. Let  $\bm{u} = \frac{\bm{\mu}}{\Vert\bm{\mu}\Vert}$ be the direction of the mean vector so that $\bm{\mu}=c_0\bm{u}$ where $c_0\in \mathbb{R}^{+}$ and $\bm{u}$ lies on a unit sphere; $c_0$ can be interpreted as the radius of the sphere on which the mean vector lie. Our structured covariance model is
\begin{align}
\label{SigStruc}
\bm{\Sigma}&=\bm{\Sigma}(\bm{u},\lambda_1,\dots,\lambda_{p-1}) = \bm{P}(\bm{u}) \bm{D} \bm{P}^{\top}(\bm{u})
\end{align}
where $\bm{D} = \text{ diag }(1,\lambda_1,\dots,\lambda_{p-1}) = \text{ diag }(1,\bm{\lambda})$ is the matrix of eigenvalues, and for a given value of the mean direction $\bm{u}$, the orthogonal matrix of eigenvectors is $\bm{P}(\bm{u})=[\bm{u}, \bm{V}(\bm{u})]$ where $\bm{V}=\bm{V}(\bm{u})\in {R}^{p\times (p-1)}$ is a known matrix function so that orthogonality of $\bm{P}$ is ensured. A simple and prominent  example of such a  $\bm{V}$ is obtained by an application of Gram-Schmidt procedure \citep{trefethen1997numerical}  to the set $\{\bm{u},\bm{e}_1,\bm{e}_2,\dots,\bm{e}_{p-1}\}$ where $\bm{e}_i, i=1,2,\dots,p-1$ denote the canonical basis of $\mathcal{R}^p$ and $u_p\neq 0$. 
 With $\bm{V}$ assumed known in (\ref{SigStruc}), the unknown parameters then are $(\bm{\mu}, \bm{\lambda})$, the vectors of mean and  the  $(p-1)$ eigenvalues, so that the  number of parameters drastically reduces from quadratic  to linear i.e. $(2p-1)$ in the dimension. Moreover,  the first constraint is automatically satisfied, since using $\bm{\mu}\perp \bm{V}_i(\bm{u})=\bm{V}_i$ (columns of the $\bm{V}$ matrix) for $i=1,2,\dots,(p-1)$, it follows that

\begin{align}
\bm{\Sigma}\bm{\mu} &=  \bm{P}(\bm{u}) \bm{\Lambda} \bm{P}^{\top}(\bm{u}) \bm{\mu}= \sum_{i=1}^{p-1} \lambda_i \bm{V}_i\bm{V}^{\top}_i\bm{\mu} + \bm{u}\bm{u}^{\top}\bm{\mu}= c_0\bm{u}\bm{u}^{\top}\bm{u} =\bm{\mu}.
\end{align}
Our approximate MLE of the parameters in Section 2 provides positive estimates for the eigenvalues and hence guarantees the positive-definiteness
of the estimated structured covariance matrix.

  Next, we will provide two concrete examples to elucidate the idea behind the model and its components, particularly the eigenvectors as functions of the mean vector.

\noindent \textbf{Example 1:} Consider the case of equal means $\bm{\mu} = c\mathbb{1}$, then using Gram-Schmidt procedure \citep{trefethen1997numerical}, it follows that $\bm{P}(\bm{u})$ is of the  form:
\begin{align}
\bm{P}(\bm{u})&=\begin{bmatrix}
\mathbb{1}_p & \bm{z}^{(p)}_p & \bm{z}^{(p)}_{p-1} & \bm{z}^{(p)}_{p-2} & \dots & \bm{z}^{(p)}_{2}
\end{bmatrix}
\end{align}
where
\begin{align}
\bm{z}^{(p)}_{s} &= \left(0,\dots,0,\bm{z}^{(s)}_s\right)^{\top} \quad\quad\quad \forall s\in\left\{2,3\dots,p\right\}\nonumber\\
\bm{z}^{(s)}_s &= \left(\frac{s-1}{\sqrt{s(s-1)}},\frac{-1}{\sqrt{s(s-1)}},\dots,\frac{-1}{\sqrt{s(s-1)}}\right)^{\top}
\end{align}
 and $\mathbb{1}_p = \left(\frac{1}{\sqrt{p}},\frac{1}{\sqrt{p}},\dots,\frac{1}{\sqrt{p}}\right)$. The $c$ drops due to normalization.

\noindent \textbf{Example 2:} As a generalization  of Example 1, take $$\bm{\mu} = (\mu_1,\mu_2,\mu_3,\dots, \mu_3)^{\top}$$ where the first three entries of the mean vector are different and rest of them are the same as the third entry of the vector. Then, using the construction method described in the paragraph following (\ref{SigStruc}), the $\bm{P}$ matrix takes the  form:

\begin{align}
\bm{P}(\bm{u}) = \begin{bmatrix}
\bm{u} & \bm{w}_1 & \bm{w}_2 & z_{p-2}^{(p)} &\dots z_{2}^{(p)}
\end{bmatrix}
\end{align}
 where $z_s^{(p)}$'s are same as in Example 1, and
 \begin{align}
 \label{Ex2EV}
 \bm{w}_1 &= \left(\mu_0^2, -\mu_1\mu_2,-\mu_1\mu_3,\dots,-\mu_1\mu_3\right)^{\top}/\left(\mu_0\Vert\mu\Vert\right)\nonumber\\
 \bm{w}_2 &= \left(0, (p-2)\mu_3, -\mu_2,\dots,-\mu_2\right)^{\top}/\left((p-2)\mu_0\right),
 \end{align}
  where $\mu_0^2 = \mu_2^2 + (p-2)\mu_3^2$. When $p=3$, $\bm{w}_1$ and $\bm{w}_2$ reduce to the  $\tilde{\xi}_1$ and $\tilde{\xi}_2$ in  \citet[\S 2.3]{paine2018elliptically}, so that this example  is a generalization of their construction  to the case of $p> 3$ or more general spherical data.

                      Spherical data arise in many scientific disciplines like shape analysis, geology, meteorology (e.g. \cite{mardia2000directional}), text analysis (e.g. \cite{hamsici2007spherical}), etc. \cite{paine2020spherical} study spherical regression using their elliptically symmetric angular Gaussian distribution. Since our constraint is similar without the spherical nature of the distribution, we can always project them in the corresponding $\mathbb{R}^p$ \citep[\S 3.5.6]{mardia2000directional} and fit a normal model with the constraint (\ref{ESAGCon}). This is helpful in developing multivariate techniques like discriminant analysis, clustering, etc. for spherical data. We have applied it on earth's historic magnetic pole data by \cite{schmidt1976non}.

Our goal is to compute the maximum likelihood estimator and provide a Bayesian framework for estimation of the parameters in (\ref{SigStruc}). It turns out that computing the maximum
likelihood estimator is challenging due to the nonlinearity and intractability of $\bm{P}(\bm{u})$. However, the MLE of the eigenvalues and $c_0$ have closed forms as a function of the mean direction, thus making it possible to compute its profile likelihood.
At this stage, we approximate the MLE of the mean direction by first finding a lower bound for the concave profile log-likelihood function of the mean direction (for given $\bm{\lambda}$) and then
maximizing it. In the Bayesian context, the maximum aposteriori (MAP) does not have a closed form, even though the posterior distribution obtained by using a Gaussian prior
for the mean vector and inverse gamma on the eigenvalues does. These priors are suggested by extending the shrinkage inverse Wishart prior of \cite{berger2020bayesian} to the case of nonzero means. We  show in Section \ref{GibbsSampNM}  that
one can generate from the posterior distribution quite easily using Metropolis-Hastings within Gibbs sampling. But if we are only interested in point estimation (e.g. MAP), then we propose a simpler way of
computing it. The idea is to follow the calculation of the approximate MLE by providing a lower bound for the posterior and maximize it using a modified version of
Newton's method.

The paper is organized as follows: Section \ref{MLE} delves into the detailed computation of an approximate maximum likelihood estimator. Section \ref{JointBE} discusses the standard priors such as normal-inverse Wishart and normal-shrinkage inverse Wishart, motivates the normal-inverse gamma prior selection,  provide details of the MH within Gibbs sampling process for  generation from the posterior and imitates the approximate MLE calculation to propose a fast method to compute the Bayesian point estimate. Section \ref{SimSec} computes the risks of these methods of approximation through simulation and exhibits the performance of our method.  Finally in section \ref{Conc}, we discuss the key take away for this paper.

\section{MLE of $\mu$ and $\lambda$ in Model   (\ref{SigStruc})}
\label{MLE}

Let $\bm{x}_1,\bm{x}_2,\dots,\bm{x}_n$ be a sample of size $n$ from $N_p(\bm{\mu},\bm{\Sigma})$, where $\bm{\Sigma}$ is parameterized as in (\ref{SigStruc}) with $\bm{\mu} = c_0\bm{u}$ and  $\bm{X}$ is the $n\times p$ data matrix. Then, the log-likelihood can be written as

\begin{align}
l(\bm{u},c_0,\bm{\lambda}\mid\bm{X}) &\propto  -\frac{n}{2}\sum_{i=1}^{p-1} \log(\lambda_i) - \frac{1}{2}\left[\sum_{j=1}^n(c_0\bm{u} - \bm{x}_i)^{\top}\bm{\Sigma}^{-1}(c_0\bm{u} - \bm{x}_i)\right]\label{loglik1:1}\\
 &\propto -\frac{n}{2}\sum_{i=1}^{p-1} \log(\lambda_i) - \frac{1}{2} \text{Tr}\left[\bm{D}^{-1}\bm{B}(\bm{X},c_0,\bm{u})\right]\nonumber\\
&\propto  -\frac{n}{2}\sum_{i=1}^{p-1} \log(\lambda_i) -\frac{1}{2} \left[\bm{B}(\bm{X},c_0,\bm{u})_{11} + \sum_{i=1}^{p-1} \frac{\bm{B}(\bm{X},c_0,\bm{u})_{(i+1)(i+1)}}{\lambda_i}\right]\label{loglik1:2},
\end{align}
where $\bm{B}(\bm{X},c_0,\bm{u}) = \bm{P}^{\top}(\bm{u})\bm{A}(c_0\bm{u})\bm{P}(\bm{u})$ and $\bm{A}(\bm{\mu})=\sum_{i=1}^n (\bm{x}_i - \bm{\mu})(\bm{x}_i - \bm{\mu})^{\top}$.

Evidently, maximization of the log-likelihood function with respect to the mean direction is more challenging than those with respect to $c_0$ and $\bm{\lambda}$. This suggests and we propose approximating the MLE of the parameters using the following three (non-iterative) steps:
\begin{enumerate}
\item Differentiate  the log-likelihood function to obtain the MLE of $c_0$ and $\lambda_i$'s as a function of the mean direction vector $\bm{u}$.
\item Compute the profile log-likelihood.
\item Find a lower bound for the  profile log-likelihood  and maximize it to obtain the approximate MLE of the mean direction $\bm{u}$. Use the latter   to compute the approximate MLE of $c_0$ and the eigenvalues.

\end{enumerate}

Next, we provide further details about  implementing these three steps.

\noindent \textbf{Step 1:}
 For a given $\bm{u}$, differentiating the log-likelihood in (\ref{loglik1:1}) with respect to $c_0$ \citep[\S 4.2.9]{bibby1979multivariate}  and in (\ref{loglik1:2}) with respect to $\lambda_i$, we obtain
\begin{align}
\label{lambEst}
\widehat{c}_0 = \frac{\bm{u}^{\top}\bm{P}(\bm{u})\bm{D}^{-1}\bm{P}^{\top}(\bm{u})\bar{\bm{x}}}{\bm{u}^{\top}\bm{P}(\bm{u})\bm{D}^{-1}\bm{P}^{\top}(\bm{u})\bm{u}}, \quad &\text{ and }\quad
\widehat{\lambda}_i = \frac{\bm{B}(\bm{X},c_0,\bm{u})_{(i+1)(i+1)}}{n}.
\end{align}
\noindent The expression for the $\widehat{c}_0$ in (\ref{lambEst}) reduces to
\begin{align}
\label{c0Est}
\widehat{c}_0 &= \frac{\bm{e}^{\top}_1\bm{D}^{-1}\bm{P}^{\top}(\bm{u})\bar{\bm{x}}}{\bm{e}_1^{\top}\bm{D}^{-1}\bm{e}_1}=\frac{\frac{1}{\lambda_1}\bm{e}^{\top}_1\bm{P}^{\top}(\bm{u})\bar{\bm{x}}}{\frac{1}{\lambda_1}} = \bm{u}^{\top}\bar{\bm{x}},
\end{align}
and note that the estimates of the eigenvalues are always positive.

\noindent \textbf{Step 2:} Substituting for $\widehat{c}_0$ and $\widehat{\lambda}_i$ in the log-likelihood (\ref{loglik1:2}), the profile likelihood of the mean direction $\bm{u}$ turns out to be

\begin{align}
\label{proflik1}
l(\bm{u},\widehat{c}_0,\widehat{\bm{\lambda}}\mid\bm{X}) &\propto -\frac{n}{2}\sum_{i=1}^{p-1} \log\left(\frac{\bm{B}(\bm{X},\widehat{c}_0,\bm{u})_{(i+1)(i+1)}}{n}\right) -\frac{1}{2}\left[\bm{B}(\bm{X},\widehat{c}_0,\bm{u})_{11} + n(p-1)\right]
\end{align}
 Denoting the columns of $\bm{V}$ by $\bm{V}_i$ for $i=1,2,\dots,(p-1)$, then the diagonal entries of the $\bm{B}(\bm{X},\widehat{c}_0,\bm{u})$ are
\begin{align*}
&\left\{\bm{B}(\bm{X},\widehat{c}_0,\bm{u})_{11},\bm{B}(\bm{X},\widehat{c}_0,\bm{u})_{22},\dots,\bm{B}(\bm{X},\widehat{c}_0,\bm{u})_{p,p}\right\}\\
&=\left\{\bm{u}^{\top}\bm{A}(\widehat{c}_0\bm{u})\bm{u}, \bm{V}^{\top}_1\bm{A}(\widehat{c}_0\bm{u})\bm{V}_1,\dots,\bm{V}^{\top}_{p-1}\bm{A}(\widehat{c}_0\bm{u})\bm{V}_{p-1}\right\}.
\end{align*}
Since  $\bm{u}\perp\bm{V}_i$ for $i=1,2,\dots,{p-1}$, one may further simplify the diagonal entries of $\bm{B}(\bm{X},\widehat{c}_0,\bm{u})$ to the following:
\begin{align}
\label{BXmu}
\bm{V}_i^{\top} \bm{A}(\widehat{c}_0\bm{u}) \bm{V}_i &= \bm{V}_i^{\top} \left[\bm{A}(\bm{0}) - n\widehat{c}_0\bar{\bm{x}}\bm{u}^{\top} - n\widehat{c}_0\bm{u}\bar{\bm{x}}^{\top}+ n\widehat{c}_0^2\bm{u}\bm{u}^{\top}\right] \bm{V}_i = \bm{V}_i^{\top}\bm{A}(\bm{0})\bm{V}_i\nonumber\\
\bm{u}^{\top}\bm{A}(\widehat{c}_0\bm{u})\bm{u} &= \bm{u}^{\top}\bm{A}(\bar{\bm{x}})\bm{u}.
\end{align}
In spite of this simplification, the profile log-likelihood is  hard to differentiate as a function of $\bm{u}$, in general, when we are not assuming any specific form for the matrix $\bm{V}$.

\vspace*{2mm}
\noindent \textbf{Step 3:} We find  a workable lower bound for (\ref{proflik1}) and maximize it with respect to the mean direction to obtain an approximate MLE for $\bm{u}$. Alternatively, this amounts to assuming that the first sum in the profile likelihood (2.7) is constant. The expression in equation (\ref{BXmu}) is a quadratic form of the matrix $\bm{A(0)}=\sum_{j=1}^n\bm{x}_j\bm{x}_j^{\top}$ and can be bounded by its largest eigenvalue \citep[\S 1f.2.1]{rao1973linear}. Consequently, using  (\ref{BXmu}) and that columns of $\bm{V}$ are orthonormal simplify the profile log-likelihood (\ref{proflik1}) and  we arrive at the following  lower bound:

 \begin{align}
 \label{proflikineq}
 & l(\bm{u},\widehat{c}_0,\bm{\lambda}\mid\bm{X})\propto -\frac{n}{2}\sum_{i=1}^{p-1} \log\left[\frac{\bm{V}_i^{\top}\bm{A}(\bm{0})\bm{V}_i}{n}\right] -\frac{1}{2}\left[ \bm{u}^{\top}\bm{A}(\bar{\bm{x}})\bm{u} + n(p-1) \right] \nonumber\\
 &\geq  -\frac{n(p-1)}{2} \log\left[\frac{\lambda_1\left\{\bm{A}(\bm{0})\right\}}{n}\right]-\frac{1}{2}\left[ \bm{u}^{\top}\bm{A}(\bar{\bm{x}})\bm{u} + n(p-1) \right]= h(\bm{u})
  \end{align}

  Evidently, the lower bound denoted by $h(\bm{u})$  is maximized by minimizing the quadratic expression inside the second bracket. Since $\bm{u}\neq 0$ and  $h$ is a quadratic function of $\bm{u}$, the maximum occurs when $\bm{u}$ is the eigenvector corresponding to the smallest eigenvalue of the matrix $\bm{A}(\bar{\bm{x}})$. As such the approximate MLE of $\bm{u}$ is unique only up to a sign.

  Even though the computed mean direction $\widehat{\bm{u}}$ is not the exact MLE, it is still a good approximation as confirmed by the simulation results  in Table \ref{tab:mlemapNR} of Section \ref{SimSec}. The approximate maximum likelihood estimate of the eigenvalues and the constant $c_0$ are obtained by plugging in the approximate MLE of $\bm{u}$ in (\ref{lambEst}) and (\ref{c0Est}). Fortunately, this idea of approximatiing  MLE can also be replicated in our posterior MAP estimator approximation developed in the next section, but the maximization of the lower bound is not as straightforward and requires employing a version of Newton's iterative method.


\section{Bayesian Estimation of $(\mu,\lambda)$ in   (\ref{SigStruc})}
\label{JointBE}

In this section we develop a Bayesian methodology for estimating the parameters using two sets of priors reviewed in
the next two subsections.

\subsection{Normal-Inverse Wishart Priors:}
\label{NIWPrior}

 For the parameters of a multivariate normal distribution the most popular prior is normal-inverse Wishart described in \citet[Section 3.6]{gelman2013bayesian} and \cite{barnard2000modeling}  with the hyperparameters $(\bm{\mu}_0,\kappa_0;\nu_0,\bm{\Lambda}_0)$:

\begin{align*}
\bm{\mu}\mid\bm{\Sigma} \sim N_p(\bm{\mu}_0,\bm{\Sigma}/\kappa_0),\qquad \bm{\Sigma} \sim \pi_{IW}(\nu_0, \bm{\Lambda}^{-1}_0),
\end{align*}

\noindent where $\nu_0$ and $\bm{\Lambda}_0$  are the degrees of freedom and the scale matrix. The suggested values for the hyperparameters
 in \cite{gelman2006data} are $\bm{\Lambda}_0=\bm I_p$ and $\nu_0=p+1$. The remaining hyperparameters are the prior
  mean, $\bm{\mu}_0$ and $\kappa_0$ on the $\bm{\Sigma}$ scale. Due to conjugacy, the posterior density has the following parameters:

\begin{align}
\label{PostPar}
\bm{\mu}_n &= \frac{\kappa_0}{\kappa_0+n} \bm{\mu}_0 +\frac{n}{\kappa_0+n} \bar{\bm x}\nonumber,\\
\bm{\Lambda}_n &= \bm{\Lambda}_0 + \bm A + \frac{n\kappa_0}{\kappa_0+n}(\bar{\bm x}-\bm{\mu}_0)(\bar{\bm x}-\bm{\mu}_0)^{\top}.
\end{align}
 As a point estimator posterior mode or the maximum aposteriori probability (MAP) estimator for $\bm{\Sigma}$ is generally the most popular
  choice  \citep[\S 5.2.1]{murphy2012machine}. The posterior  marginal distribution of the mean parameter $\bm{\mu}$ is  multivariate $t_{\nu_n-p+1}\left(\bm{\mu}_n,\frac{\bm{\Lambda}_n}{\kappa_n(\nu_n-p+1)}\right)$ with the MAP estimator $\bm{\mu}_n$, and the MAP estimator for $\bm{\Sigma}$ is  given by
\begin{align*}
\widehat{\bm{\Sigma}}_{map} &= \frac{\bm{\Lambda}_n}{\nu_n+p+2},
\end{align*}
where
  $\nu_n=\nu_0+n$ (proof is shown in appendix \ref{MAPIWPr}).

\subsection{Shrinkage Inverse Wishart Priors}
\label{SIWPrior}
The most natural prior for mean and covariance i.e. normal-inverse Wishart reviewed above is known to overdisperse the eigenvalues of the posterior  covariance estimator. In fact, \cite{yang1994estimation} describes that the normal-inverse Wishart prior has the term $\prod_{i<j} (\lambda_i - \lambda_j)$ in the density where $\lambda_i$'s are the ordered eigenvalues of $\bm{\Sigma}$, thus forcing the eigenvalues apart and increasing the variability \citep{berger2020bayesian}.  This is one major motivation for using the normal-shrinkage inverse Wishart prior.\par 

The shrinkage-inverse Wishart (SIW) prior has the same density as inverse Wishart prior save the extra term $\prod_{i<j} (\lambda_i - \lambda_j)$ in the denominator:
\begin{align}
\pi_{SIW}(\bm{\Sigma}\mid \nu_0, b , \bm{\Lambda}^{-1}_0) & \propto
\mid\bm{\Sigma}\mid^{-\frac{\nu_0+p+1}{2}}\frac{\exp{\left[-\frac{1}{2}Tr(\bm{\Lambda}_0\bm{\Sigma}^{-1})\right]}}{\prod_{i<j} (\lambda_i - \lambda_j)^b},
\end{align}
where  $\nu_0$ is a real constant, $b\in [0,1]$ and $\bm{\Lambda}_0$ is a positive semi-definite matrix. It is interesting  to notice  that $b=0$ corresponds to some common priors like inverse Wishart, reference, Jefferey's etc. which also contain the term $\prod_{i<j} (\lambda_i - \lambda_j)$
, see  \cite{berger2020bayesian}. The posterior density for $(\bm{D},\bm{P})$,

   using the one-to-one transformation from $\bm{\Sigma} $ to $\bm{D} = \text{diag}(\lambda_1,\dots,\lambda_p)$ and the orthogonal eigenvector matrix $\bm{P}$, turns out to be

\begin{align}
\label{DistPDSIW}
\pi_{SIW}(\bm{D},\bm{P}\mid \nu_0, b , \bm{\Lambda}^{-1}_0) & \propto
\mid\bm{\Sigma}\mid^{-\frac{\nu_0+p+1}{2}}\frac{\exp{\left[-\frac{1}{2}Tr(\bm{\Lambda}_0\bm{P}\bm{D}^{-1}\bm{P}^{\top})\right]}}{\prod_{i<j} (\lambda_i - \lambda_j)^{(b-1)}} \mathbb{1}_{\lambda_1\geq \lambda_2\geq \dots \lambda_p}.
\end{align}
Note that the posterior is  zero whenever the eigenvalues are close together so that effectively it forces the eigenvalues apart. Moreover, when $b=1$, the term in question goes away retaining the conjugacy property, even with the additional normal prior on the nonzero mean (Proof is shown in the Appendix \ref{NSIWConPr}) and the same posterior parameters as before.
 The next important consequence of choosing $b=1$, from (\ref{DistPDSIW}), is that the conditional distribution of the eigenvalues given the orthogonal matrix $\bm{P}$ is an ordered inverse gamma distribution \cite[\S 3.2]{berger2020bayesian}. This plays a central role in developing a computational scheme for computing the MAP.

   Recall that the structured covariance model  in (\ref{SigStruc}) has its orthogonal matrix $\bm{P}$ determined by a non-zero mean vector. This requires working with nonzero mean vectors in contrast to the mean zero framework in \cite{berger2020bayesian}, while maintaining the conditional distribution of the eigenvalues given the mean vector as an inverse gamma distribution, see also \cite{berger2020bayesian}. This motivates us to select an inverse gamma prior for the eigenvalues as in  (\ref{NMPrior}) which, fortunately, in turn implies that the conditional distribution of eigenvalues is the inverse gamma (see section \ref{GibbsSampNM}).

\subsection{ The Mean-Eigenvalue Priors}
\label{BayesNPNM}

 A convenient prior distribution on the mean vector is the multivariate normal and the inverse gamma on the ordered eigenvalues. The latter prior comes naturally from the shrinkage inverse Wishart prior of \cite[\S 3.2]{berger2020bayesian} as the conditional distribution of eigenvalues given the matrix of eigenvectors (see section \ref{SIWPrior} for detailed discussion), see also  \cite{hoff2009hierarchical} (section 3.3). More specifically, our proposed prior is:

\begin{align}
\label{NMPrior}
\bm{\mu}\mid \bm{D} \sim N_p(\bm{\mu}_0, \bm{D}/\kappa_0) &\text{ and }  \bm{\lambda}_i  \sim  \text{Inverse-gamma}(a-1,c_i/2),
\end{align}

\noindent where  $a, c_i$'s are the hyperparameters. Setting $\bm{H}_0 = \text{diag}(1,c_1,c_2,\dots,c_{p-1})$, then the posterior distribution has the form
\begin{align}
\label{posDen}
p(\bm{\mu},\bm{\lambda} \mid \bm{X})  &\propto   \left(\prod_{i=1}^{p-1} \lambda_i\right)^{-\frac{n+1+2a}{2}} \exp\left[-\frac{1}{2}Tr\left\{\bm{D}^{-1} \bm{H}_N\right\}\right]
\end{align}
    with $\bm{H}_N = \bm{B}(\bm{X},\bm{\mu}) + \kappa_0 (\bm{\mu}-\bm{\mu}_0)(\bm{\mu} - \bm{\mu}_0)^{\top} + \bm{H}_0$. Here $\bm{B}(\bm{X},\bm{\mu})$ has the same expression as $\bm{B}(\bm{X},c_0,\bm{u})$ which appeared in Section 2.

     One needs to generate from this posterior distribution when computing the Bayesian point estimates, credible intervals and various other quantities. We show that Metropolis-Hastings within Gibbs sampling is possible here and suitable for our goals \citep[\S 1.12.10]{brooks2011handbook}.

\subsection{Gibbs Sampling}
\label{GibbsSampNM}
  The Gibbs sampler \citep{gelfand1990sampling} method is a numerical technique for sampling from the joint posterior distribution. Given an initial vector, the Gibbs sampling proceeds by sampling from the conditional posterior distribution. More generally, if the full conditional posterior distribution in any Gibbs step is of a non-standard form, using MH (Metropolis-Hastings) step is convenient \citep{brooks2011handbook}. The technique is known as Metropolis-Hastings within Gibbs Sampling or alternatively as single-component Metropolis-Hastings sampling as follows:

\begin{enumerate}
\item Since  $p(\bm{D}\mid \bm{\mu},\bm{X}) \propto p(\bm{\mu},\bm{\lambda} \mid \bm{X}) $  has an Inverse-Gamma distribution, the orderd  eigenvalues are generated from independent Inverse Gamma $\left(\frac{n+2a-1}{2} , c^*_i/2\right)$  where $c^*_i$ is the i-th diagonal entry of $\bm{H}_N$ matrix.
\item Generate from $p(\bm{\mu}\mid \bm{D}, \bm{X})$ using  the  MH algorithm with a  proposal distribution $q$  selected as

\begin{align*}
q\left(. \mid \bm{\mu}^{(i-1)}\right) = N_p\left(\bm{\mu}^{(i-1)}, \frac{1}{n}\bm{P}\left(\bm{\mu}^{(i-1}\right)\bm{D}\bm{P}^{\top}\left(\bm{\mu}^{(i-1)}\right)\right).
\end{align*}
\end{enumerate}

\noindent More concretely, the steps of MH within Gibbs algorithm in our context are as follows:

 \begin{breakablealgorithm}
 \caption{MH within Gibbs for generating samples from the posterior distribution}
 \begin{algorithmic}[1]
\State Start with $\bm{\mu}^{(0)} = \bar{\bm{x}}$.
\State \textbf{Repeat $s$ times:} $j-th$ step
\State \quad Generate $\bm{\lambda}^{(j)}_i \sim IG\left(\frac{n+2a-1}{2}, c^{*}_i/2\right)$ to form $\bm{D}^{(j)}$.
\State \quad Start with $\bm{\mu}^{(j-1)}$, \textbf{Repeat MH Step $l$ times:} $k-th$ step
\State \quad \quad Generate $\bm{\mu}^{*}\sim q\left(.\mid \bm{\mu}^{(k-1)}\right)$
\State \quad\quad Calculate $$r\left(\bm{\mu}^{*},\bm{\mu}^{(k-1)}\right) = \min\left\{1, \frac{p\left(\bm{\mu}^*, \bm{\lambda}^{(j)} \mid \bm{X}\right) q\left(\bm{\mu}^{(k-1)}\mid \bm{\mu}^{*}\right) }{q\left(\bm{\mu}^*\mid \bm{\mu}^{(k-1)}\right) p\left(\bm{\mu}^{(k-1)}, \bm{\lambda}^{(j)} \mid \bm{X}\right)  }\right\}$$
\State \quad \quad Generate $u\sim U(0,1)$
\State \quad \quad Set $\bm{\mu}^{(k)}=\bm{\mu}^{*}$ if $u<r\left(\bm{\mu}^{*},\bm{\mu}^{(k-1)}\right)$ else  $\bm{\mu}^{(k)}=\bm{\mu}^{(k-1)}$
\State \quad Set $\bm{\mu}^{(j)} = \bm{\mu}^{(l)}$
\State Collect $(\bm{\mu}^{(j)},\bm{\lambda}^{(j)})$ for $j=1,2,\dots, s$
\State \textbf{End}
\end{algorithmic}
\end{breakablealgorithm}

 An interesting observation is that the initial value of $\bm{\lambda}$ is irrelevant. In the MCMC, the chain for the eigenvalues depend on the previous iteration only through the generation of mean vector. This is happening because the parameters for the conditional distribution of the eigenvalues are $(n+2a-1)/2$ and $c^*_i/2$ where $c^*_i$'s are the diagonal entries of the $\bm{H}_N$ matrix and hence a function of the mean vector.

 To illustrate the algorithm, we have generated 100 pairs of $(\bm{\mu},\bm{\lambda})$ from the posterior distribution by MH within Gibbs sampling.  We find out the posterior maximizing $(\bm{\mu},\bm{\lambda})$ pair by exhaustive computation of posterior density at each pair. This is possible because the posterior density has a closed form barring the normalizing constants. We take the value of the mean vector from the posterior maximizing pair, set it as our MAP estimate for the mean vector discarding the corresponding $\bm{\lambda}$ and make a final update on $\bm{\lambda}$. Given the estimate of the mean vector, the posterior conditional distribution of $\bm{\lambda}$ is inverse gamma and has a known mode. Using this, we can calculate the best eigenvalue vector $\widehat{\bm{\lambda}}$ to be $\widehat{\lambda}_i = \frac{c_i^*}{n+1+2a}$ for $i=1,2,\dots, (p-1)$. The newly calculated value $\widehat{\lambda}$, accompanied by the MAP estimate of $\bm{\mu}$ constitute our MAP estimator  $\left(\widehat{\mu},\widehat{\lambda}\right)$. The performance of such an estimator is assessed through a simulation study (see Section \ref{GibbSim}) and Table \ref{tab:Gibbstable} of the simulation Section 4.

\subsection{Approximation of MAP through a Lower Bound}
\label{MAPLB}
  The Gibbs sampling is computationally challenging and the time complexity increases exponentially with the dimensions. We  follow the steps Section 2 to approximate the MAP estimator using a lower bound for the log posterior density. The main difference is in the  maximization step of the lower bound as a function of $\bm{u}$ where  it is not as straightforward as the MLE case and does not have a closed-form. So, we resort to an iterative modified Newton-Raphson algorithm.
\vspace*{2mm}
     \noindent Using $\bm{\mu} = c_0\bm{u}$ where $c_0\in \mathbb{R}$ as before we arrive at
\begin{align}
\log p(\bm{u}, c_0, \bm{\lambda}\mid\bm{X}) & \propto -t\sum_{i=1}^{p-1} \log \lambda_i - \frac{1}{2}\left[ f(c_0) + \text{ Tr }(\bm{D}^{-1}\bm{H}_0) \right]\\
&\propto -t\sum_{i=1}^{p-1} \log \lambda_i -\frac{1}{2}\left[\left(\bm{H}_N\right)_{11}+\sum_{i=1}^{p-1} \frac{\left(\bm{H}_N\right)_{i+1,i+1}}{\lambda_i}\right]
\end{align}
  where $t=\frac{n+1+2a}{2}$ and $f(c_0) = \sum_{i=1}^n(\bm{x}_i-c_0\bm{u})^{\top}\bm{\Sigma}^{-1}(\bm{x}_i-c_0\bm{u}) + \kappa_0 (c_0\bm{u}-\bm{\mu}_0)^{\top}\bm{D}^{-1} (c_0\bm{u}-\bm{\mu}_0).$  Following  similar calculations as in (\ref{lambEst}), the estimates of $c_0$ and the eigenvalues in terms of the mean direction are as follows:
\begin{align}
\label{c0lambMAP}
\widehat{c}_0 = \frac{n\bm{u}^{\top}\bar{\bm{x}} + \kappa_0\bm{u}^{\top}\bm{D}^{-1}\bm{\mu}_0}{n+\kappa_0\bm{u}^{\top}\bm{D}^{-1}\bm{u}} ,\quad &\quad
\widehat{\lambda}_i = \frac{\left(\bm{H}_N\right)_{i+1,i+1}}{2t}
\end{align}
 where $\bm{H}_N = \bm{B}(\bm{X},c_0,\bm{u}) + \kappa_0 (c_0\bm{u} - \bm{\mu}_0)^{\top}\bm{D}^{-1}(c_0\bm{u} - \bm{\mu}_0) + \bm{H}_0$. Here the expression of $\widehat{c}_0$ and $\widehat{\lambda}_i$ are intertwined. So we are doing the following calculation with the aim to propose an iterative algorithm for the MAP estimators. We will use equation (\ref{c0lambMAP}) and (\ref{NRup}) as our updating equation.

  Using the expression of $\widehat{\lambda}_i$ from equation (\ref{c0lambMAP}) in the log posterior likelihood function we obtain the following  analogue of the profile likelihood in (\ref{proflik1}):
  \begin{align}
  \label{postproflik1}
  \log p(\bm{u}, c_0, \widehat{\bm{\lambda}}\mid\bm{X}) &\propto -t \sum_{i=1}^{p-1} \log \left(\frac{\left(\bm{H}_N\right)_{i+1,i+1}}{2t}\right) -\frac{1}{2}\left[\left(\bm{H}_N\right)_{11}+(p-1)t\right]
  \end{align}
  where the diagonal entries of $\bm{H}_N$ are
  \begin{align}
  \label{HNDiagMAP}
\left(\bm{H}_N\right)_{i+1,i+1}&=  \begin{cases}
\bm{u}^{\top}\bm{A}(c_0\bm{u})\bm{u} + \kappa_0 \left(c_0\bm{u}-\bm{\mu}_0\right)^2_1 + \left(\bm{H}_0\right)_{1,1}  \quad\quad\quad \text{ for } i=0\\
\bm{V}_{i}^{\top}\bm{A}(0)\bm{V}_{i} + \kappa_0 \left(c_0\bm{u}-\bm{\mu}_0\right)^2_{i+1} + \left(\bm{H}_0\right)_{i+1,i+1}  \quad\quad\quad \text{ for } i\neq 0 \text{ and } \forall c_0
  \end{cases}.
\end{align}
  This follows from the expression of the matrix $\bm{B}(\bm{X}, c_0, \bm{u})$ in equation in (\ref{BXmu}) and the definition of $\bm{H}_N$.
   We observe the following
\begin{align}
\left(\bm{H}_N\right)_{i+1,i+1} &\leq \lambda_1\left\{\bm{A}(\bm{0})\right\} + \kappa_0\Vert c_0\bm{u} - \bm{\mu}_0\Vert^2 + \left(\bm{H}_0\right)_{i+1,i+1} \nonumber\\
&= m_i + \kappa_0\Vert c_0\bm{u} - \bm{\mu}_0\Vert^2  \quad  \text{ for } i\neq 0 \text{ and } \forall c_0 \in \mathbb{R}.
\end{align}
For a given value of $c_0$, equation (\ref{postproflik1}), provides a lower bound $h(\bm{u})$, which we will maximize to approximate $\bm{u}$. This is equivalent to minimizing $-h(\bm{u})$. We apply Newton - Raphson algorithm \citep{lange2013elementary} to obtain an update for $\bm{u}$. The calculation of derivatives are shown below.
  \begin{align}
  &\log p(\bm{u}, c_0, \widehat{\bm{\lambda}}\mid\bm{X}) \geq h(\bm{u})\nonumber,\\
& -h(\bm{u})= t\sum_{i=1}^{p-1} \log\left[\frac{m_i+\kappa_0 \Vert c_0\bm{u}-\bm{\mu}_0\Vert^2}{2t}\right] + \frac{1}{2}\left[\bm{u}^{\top}\bm{A}(c_0\bm{u})\bm{u} + \kappa_0 \Vert c_0\bm{u}-\bm{\mu}_0\Vert^2_1 + m_1 \right]\label{h},\\
&-\nabla h(\bm{u}) = t \sum_{i=1}^{p-1} \frac{4t\kappa_0c_0(c_0\bm{u} - \bm{\mu}_0) }{m_i+\kappa_0 \Vert c_0\bm{u}-\bm{\mu}_0\Vert^2} + \left[\bm{A}(0)\bm{u} - c_0 n \bm{\bar{x}} +\kappa_0c_0(c_0\bm{u} - \bm{\mu}_0)\right]\label{hder1},\\
&-\nabla^2 h(\bm{u}) =  t\sum_{i=1}^{p-1} \frac{4t\kappa_0 c_0\left[(m_i+\kappa_0 \Vert c_0\bm{u}-\bm{\mu}_0\Vert^2)\bm{I} - 2(c_0\bm{u} - \bm{mu}_0)(c_0\bm{u} - \bm{\mu}_0)^{\top}\right]}{\left\{m_i+\kappa_0 \Vert c_0\bm{u}-\bm{\mu}_0\Vert^2\right\}^2}\nonumber,\\
&\quad \quad \quad \quad \quad \quad + \left[\bm{A}(0) + \kappa_0 c^2_0(c_0\bm{u} - \bm{\mu}_0)(c_0\bm{u} - \bm{\mu}_0)^{\top}\right]\label{hder2}
  \end{align}
 Thus, the updating equation for the mean direction is
  \begin{align}
  \label{NRup}
  \bm{u}^{(k+1)} = \bm{u}^{(k)} - \left[\nabla^2 h(\bm{u}) \right]^{-1}\nabla h(\bm{u})
  \end{align}

Due to the concavity of the lower bound, Newton's method is  an appealing choice in our case. However there are two potential problems with Newton's method \citep[\S 10.3]{lange2013elementary}. First, it may be computationally expensive to invert the second derivative matrix in each step. Second, the Newton's method is not really a ascent algorithm in the sense that $h\left(\bm{u}^{(k+1)}\right)>h\left(\bm{u}^{(k)}\right)$ for a concave function. The second problem can be remedied by modifying the increment such that it is a partial step in the ascent direction. Let us denote:
  \begin{align}
  \bm{v} =  \left[\nabla^2 h(\bm{u})\right]^{-1}\nabla h(\bm{u})
  \end{align}
The idea is to take a sufficiently short increment in the direction of $\bm{v}$.  If $\left[\bm{u}^{(k)} - \alpha\bm{v}\right]$ shows increment in $h(\bm{u})$ value, we update our mean vector in the iteration, otherwise we look at $\left[\bm{u}^{(k)} - \alpha^j\bm{v}\right]$ for $j=1,2,\dots$ until we observe an increment. Due to the good performance of the MLE, we use the MLE of $c_0$ and $\bm{u}$ as our initial values which make the convergence fast. Based on (\ref{c0lambMAP}) and (\ref{NRup}), an iterative algorithm for computing the approximate  MAP is summarized in the following algorithm:
 \begin{breakablealgorithm}
 \caption{MAP Approximation}
 \begin{algorithmic}[1]
 \State \textbf{Initialize:} Start with $c^{(0)}_0=$ MLE of $c_0$, $\bm{u}^{(0)} =$ MLE of $\bm{u}$ and $\lambda^{(0)}_i = \frac{\bm{H}_N\left(c^{(0)},\bm{u}^{(0)}\right)_{i+1,i+1}}{2t}$
 \State \textbf{For }   $k\rightarrow (k+1)$
 \State $\quad $  Update $c^{(k+1)}$ and $\lambda^{(k+1)}_i$ from equation (\ref{c0lambMAP})
    
\State $\quad $ \textbf{For} $j \rightarrow (j+1)$\\
$\quad \quad $ $\bm{u}^{(j)} = \bm{u}^{(k)}$ and  $\bm{v}^{(j)} = \left[\nabla^2 h(\bm{u}^{(j)}) \right]^{-1}\nabla h(\bm{u}^{(j)})$
 \State $\quad \quad $\textbf{ For } $l \rightarrow (l+1)$\\
$\quad \quad \quad $ $\bm{u}^{(j+1)} = \bm{u}^{(j)} - \alpha^{l}\bm{v}^{(j)} $\\
$\quad \quad \quad$ \textbf{If } $h(\bm{u}^{(j+1)}) > h(\bm{u}^{(j)})$ \\
$\quad \quad \quad \quad$  Accept the update of $\bm{u}$\\
$\quad \quad \quad \quad $ \textbf{Break }\\
$\quad \quad \quad $ \textbf{Else } Change $\alpha^l\rightarrow\alpha^{l+1}$\\
$\quad \quad $ \textbf{If } $\Vert \bm{u}^{(j+1)} - \bm{u}^{(j)}\Vert < \epsilon$\\
$\quad \quad \quad$ $\bm{u}^{(k+1)} = \bm{u}^{(j+1)}$\\
 $\quad \quad \quad$ \textbf{Break}
\State $\quad $ \textbf{If } $h(\bm{u}^{k+1}) < h(\bm{u}^{(k)})$
\State $\quad $ \textbf{Break}
\State Get the final value of $c_0$, $\bm{\lambda}$ and $\bm{u}$.
\State\textbf{End}
 \end{algorithmic}
 \end{breakablealgorithm}
 This algorithm produces the MAP approximation of the eigenvalues and the mean vector. The continuity of the estimate of the covariance matrix, shown in Lemma \ref{contlem}, ensures that the covariance estimate will converge with the convergence of the mean direction vector $\bm{u}$.

  \begin{lemma}
  \label{contlem}
  If $\bm{V}(\bm{u})$, described after  (\ref{SigStruc}) is continuous, then the approximate MAP of the covariance matrix $\bm{\Sigma}$ is also a continuous function of $\bm{u}$.
  \end{lemma}
  \begin{Proof*}
   If $\bm{V}(\bm{u})$ is a continuous function of $\bm{u}$, then each column of $\bm{V}$, $\bm{V}_j$ is also continuous. This implies diagonal entries of $\bm{H}_N$ and in turn the MAP estimate of the eigenvalues are also a continuous function of $\bm{u}$. The claim follows since the MAP of $\bm{\Sigma}$ given by
   \begin{align}
   \widehat{\bm{\Sigma}} &= \bm{u}\bm{u}^{\top} + \sum_{i=1}^{p-1} \widehat{\lambda}_i \bm{V}_i\bm{V}^{\top}_i =  \bm{u}\bm{u}^{\top} + \sum_{i=1}^{p-1} \frac{\left(\bm{H}_N\right)_{i+1,i+1}}{n} \bm{V}_i\bm{V}^{\top}_i
   \end{align}
    is sum of continuous functions of $\bm{u}$.
  \end{Proof*}

\section{Simulations}

\label{SimSec}

In this section, we perform several simulations to assess the performance of:
\begin{enumerate}
\item MLE approximation through lower bound maximization (see section \ref{MLE}) of the profile likelihood function,
\item MAP estimator with normal-inverse gamma prior (see section \ref{BayesNPNM}) approximated through Gibbs sampling in section \ref{GibbsSampNM}, and
\item MAP approximation (see section \ref{MAPLB}) through posterior lower bound
\end{enumerate}
by computing their risks. Then the three estimators are compared with the MAP estimator with normal-inverse Wishart prior (see section \ref{NIWPrior}). We have used RStudio 1.3.1093 and R 4.0.3 on a 64 bit 4 Core Windows 10 laptop for all our simulations.

\subsection{ The Simulation Set up:}
We have taken the sample sizes and dimensions to be $n=50,100,300$ and $p=3,5,10$, respectively. In all cases the data generation mechanism and the risk function are kept the same. We have used Frobenius loss as our default loss function and risks are approximated by averaging the losses for 100 independent replications in each of the nine combinations of $(n,p)$.

 The parameters of the Gaussian distributions used for data generation are selected in the following way:
\begin{itemize}
\item the entries of the mean vector $\bm{\mu}$ are taken to be independent standard Gaussian variables,
 \item  the covariance matrix is generated from $\bm{\Psi}=\bm{LL}^{\top}$ where

\begin{align*}
L&= \begin{cases}
L_{ij} \sim N(0,1) \quad \text{ for } i\neq j,\\
L_{ij} \sim N(5,1) \quad \text{ for } i = j.
\end{cases}
\end{align*}
The larger diagonal entries of  $\bm L$ ensure positive-definiteness of the modified covariance matrix $\bm{\Sigma}$. \item Since such $(\bm{\mu},\bm{\Psi})$ does not necessarily satisfy conditions (\ref{ESAGCon}), the covariance matrix $\bm{\Sigma}$ is constructed from applying Lemma 4 of \cite{kundu2020mle} on $(\bm{\mu},\bm{\Psi})$.  \par
\end{itemize}

The performance of the estimators is assessed using the Frobenius (scaled $L_2$) risk as in \citet[\S 3.1]{ledoit2004well}:

$$R(\bm{\mu},\widehat{\bm{\mu}}^*)=\mathbb{E}\left[\frac{1}{p}\left\Vert\widehat{\bm{\mu}}^*-\bm{\mu}\right\Vert^2_{\mathcal{F}}\right]\qquad,\qquad R(\bm{\Sigma},\widehat{\bm{\Sigma}}^*)=\mathbb{E}\left[\frac{1}{p}\left\Vert\widehat{\bm{\Sigma}}^*-\bm{\Sigma}\right\Vert^2_{\mathcal{F}}\right],$$
     where
   $\widehat{\bm{\mu}}^*$ and $\widehat{\bm{\Sigma}}^*$ are the final estimators in each of the cases.

 \subsection{Simulation Results:}

 Here we provide the details of the simulation studies for the MLE approximation, MAP estimator computed through Gibbs sampling and using a lower bound of the posterior density. We have used normal-inverse Wishart as a yardstick to compare the risk of our estimators (MLE approximation and the two MAP's).

\subsubsection{Gibbs Sampling}
\label{GibbSim}
 The risk of the MAP estimator is computed through Gibbs sampling as described in Section \ref{GibbsSampNM}. Let us denote the initial values to be $\left(\bm{\mu}^{(0)},\bm{\lambda}^{(0)}\right)$. We have chosen $\bm{\mu}^{(0)}=\bm{\mu}_0 = \bar{\bm{x}}, \bm{H}_0 = \bm{I},\kappa_0 = 1.5$ and $a=(p+1)$. The first iteration takes in $\bm{\mu}^{(0)}$ and calculates the value of $\bm{H}_N$ as described after equation (\ref{posDen}). Using this, we can generate the value of $\bm{\lambda}^{(1)}$ from the distribution $p(\bm{D}\mid \bm{\mu},\bm{X})$ (see step 1 of Gibbs sampling in section \ref{GibbsSampNM}) which by construction makes sure that each eigenvalue is positive. We can easily see that the initial value of $\bm{\lambda}$ i.e. $\bm{\lambda}^{(0)}$ does not affect the Gibbs sampling. \par

 We have generated 100 samples from the posterior distribution using the MH within Gibbs algorithm described in section \ref{GibbsSampNM}. We repeat this experiment 100 times and approximate the risk.
 The performance of the Gibbs sampling is displayed in the following table \ref{tab:Gibbstable}. We can observe that the performance is improving as the number of observations $(n)$ increase. For low dimensional cases the risk of the MAP estimator of the covariance matrix calculated from Gibbs sampling is equivalent and better as compared to its normal-inverse Wishart counterpart. 

  A well-known problem with Gibbs sampling is that as the dimension increases the time complexity of the MAP approximation increases significantly. The acceptance rate of the MH step within Gibbs sampling is reasonable and decreases with in increment of dimension.

\begin{longtable}[c]{|c|c|c|c|c|c|}
\caption{Risk ratio of MAP approximation (from MH within Gibbs sampling) relative to the MAP of normal-inverse Wishart}
\label{tab:Gibbstable}\\
\hline
\textbf{} & \textbf{} & \multicolumn{2}{c|}{\textbf{\begin{tabular}[c]{@{}c@{}}Normal-Inverse \\ Gamma Risk\end{tabular}}} & \multirow{2}{*}{\textbf{\begin{tabular}[c]{@{}c@{}}Acc Rate\\ (in MH)\end{tabular}}} & \multirow{2}{*}{\textbf{\begin{tabular}[c]{@{}c@{}}Time  \\ (Sec.)\end{tabular}}} \\ \cline{1-4}
\textbf{n} & \textbf{p} & \textbf{Mean} & \textbf{Sigma} &  &  \\ \hline
\endfirsthead
\multicolumn{6}{c}%
{{\bfseries Table \thetable\ continued from previous page}} \\
\endhead
50 & 3 & 1.1173 & 0.8928 & 0.4135 & 980.52 \\ \hline
50 & 5 & 1.0824 & 1.0963 & 0.2822 & 2029.15 \\ \hline
50 & 10 & 1.2423 & 1.2798 & 0.1218 & 13574.59 \\ \hline
100 & 3 & 1.0542 & 0.9872 & 0.4352 & 910.92 \\ \hline
100 & 5 & 1.1677 & 1.3662 & 0.2908 & 1901.41 \\ \hline
100 & 10 & 1.1963 & 1.6229 & 0.1245 & 12183.56 \\ \hline
300 & 3 & 1.1661 & 1.2546 & 0.4287 & 1256.81 \\ \hline
300 & 5 & 1.5729 & 2.0163 & 0.2895 & 3215.97 \\ \hline
300 & 10 & 1.5533 & 2.4053 & 0.1294 & 9385 \\ \hline
\end{longtable}

\subsubsection{MLE and MAP Approximation using a Lower Bound}

 The structured covariance matrix defined in equation (\ref{SigStruc}) does not allow us to calculate the maximum likelihood estimate and the MAP estimator directly due to the intractability of the likelihood function. Hence, we approximate them through a lower bound described in Sections  \ref{MLE} and \ref{MAPLB}, respectively. The approximation of MLE performs well.
  \par

\begin{longtable}[c]{|c|c|c|c|c|c|c|}
\caption{Risk ratio of MLE and MAP approximation (through a lower bound) relative to the MAP of normal-inverse Wishart}
\label{tab:mlemapNR}\\
\hline
\textbf{} & \textbf{} & \multicolumn{2}{c|}{\textbf{\begin{tabular}[c]{@{}c@{}}MLE Approx.\\ Risk\end{tabular}}} & \multicolumn{2}{c|}{\textbf{\begin{tabular}[c]{@{}c@{}}MAP Approx. \\ Risk\end{tabular}}} & \multirow{2}{*}{\textbf{\begin{tabular}[c]{@{}c@{}}Time\\ (Sec)\end{tabular}}} \\ \cline{1-6}
\textbf{n} & \textbf{p} & \textbf{Mean} & \textbf{Sigma} & \textbf{Mean} & \textbf{Sigma} &  \\ \hline
\endfirsthead
\multicolumn{7}{c}%
{{\bfseries Table \thetable\ continued from previous page}} \\
\endhead
50 & 3 & 0.4253 & 1.1331 & 0.4253 & 1.1331 & 3.25 \\ \hline
50 & 5 & 0.6625 & 1.1528 & 0.6625 & 1.1528 & 2.59 \\ \hline
50 & 10 & 1.5009 & 1.195 & 1.5009 & 1.195 & 5.22 \\ \hline
100 & 3 & 0.3481 & 1.4383 & 0.3481 & 1.4383 & 2.52 \\ \hline
100 & 5 & 0.5342 & 1.5065 & 0.5342 & 1.5065 & 2.69 \\ \hline
100 & 10 & 1.401 & 1.63 & 1.401 & 1.63 & 2.84 \\ \hline
300 & 3 & 0.3159 & 2.3753 & 0.3159 & 2.3753 & 2.7 \\ \hline
300 & 5 & 0.5797 & 2.6429 & 0.5797 & 2.6429 & 2.95 \\ \hline
300 & 10 & 1.303 & 2.6493 & 1.303 & 2.6493 & 3.41 \\ \hline
\end{longtable}

A modified version of the Newton-Raphson method (described in the paragraph following equation (\ref{NRup})) converges relatively fast i.e. two to five iterations for MAP when initialized with the approximation of MLE of $c_0$ and $\bm{u}$. 

 The performance of these approximations (both MLE and MAP), assessed by the risk, are equivalent in dimension three and five to the Gibbs sampling technique discussed in the last section with the added benefit of being significantly faster (see the last column of table \ref{tab:mlemapNR}). For $p=10$, the performance is worse than Gibbs sampling (see Table \ref{tab:mlemapNR}).
 The usual observations like the decline of risk as $n$ increases and $p$ decreases are generally true. One interesting observation is that the performance of MLE approximation is not significantly different from the MAP approximation.

\subsubsection{An Example: Estimates of the Historic Position of
Earth’s Magnetic Pole}

 The dataset collected by \cite{schmidt1976non} contains the site mean direction estimates of the Earth's historic magnetic pole collected from 33 different sites in Tasmania. The longitude and latitudes from the data set is transformed to $X_1,X_2,\dots, X_{33}$ on a three dimensional unit sphere \citep{preston2017analysis}. The angualr gaussian distribution family is the marginal directional component of a multivariate normal distribution with ESAG distribution as a subfamily. \cite{paine2018elliptically} provided strong evidence in favor of ESAG distribution which satisfy the constraint over isotropic angular Gaussian distribution while analyzing this dataset. This inspire us to make normality assumption under the constraint similar to ESAG distribution disregarding the spherical nature of the transformed dataset. The constrained maximum likelihood estimate under the structured covariance model is:

 $$\widehat{\bm{u}}^{\top} =\begin{bmatrix}
  -0.44 & 0.32 & 0.75
 \end{bmatrix}, \quad \widehat{\bm{\Sigma}} = \begin{bmatrix}
 1.63  & -0.91 & -2.13\\
 -0.91  & 1.09 & 1.53\\
 -2.13 & 1.53 & 4.02
 \end{bmatrix} $$

 \noindent which are comparable to the maximum likelihood estimate calculated using elliptically symmetric angular Gaussian distribution with the parametrization in three dimension \citep{paine2018elliptically} i.e. $\widehat{\bm{u}}^{\top}=(-0.56,0.24,0.79)$.

\section{Discussion}
\label{Conc}
   This article demonstrates how simple ideas can be used to solve a challenging joint estimation problem under constraint (\ref{ESAGCon}).  The structured covariance model takes into account of the constraint, reduces dimensions significantly and do not suffer from the usual issue of positive definiteness. The intractability of the model is resolved through a lower bound of the profile likelihood to obtain MLE, MAP and sampling from posterior density is performed through Metropolis within Gibbs.  We have proposed a simple prior for our structured model i.e. normal on mean vector and inverse gamma on eigenvalues inspired by \cite{berger2020bayesian}.
   Although this model works nicely here for comparatively smaller dimensions, it is possible to replicate the idea in higher dimensions with some modifications. Other priors can also be explored under such a model in future.

\bibliographystyle{apalike}

\bibliography{Ref}

\begin{thebibliography}{}

\bibitem[Barnard et~al., 2000]{barnard2000modeling}
Barnard, J., McCulloch, R., and Meng, X.-L. (2000).
\newblock Modeling covariance matrices in terms of standard deviations and
  correlations, with application to shrinkage.
\newblock {\em Statistica Sinica}, pages 1281--1311.

\bibitem[Berger et~al., 2020]{berger2020bayesian}
Berger, J.~O., Sun, D., Song, C., et~al. (2020).
\newblock Bayesian analysis of the covariance matrix of a multivariate normal
  distribution with a new class of priors.
\newblock {\em Annals of Statistics}, 48(4):2381--2403.

\bibitem[Bibby et~al., 1979]{bibby1979multivariate}
Bibby, J., Kent, J., and Mardia, K. (1979).
\newblock {\em Multivariate analysis}.
\newblock Academic Press, London.

\bibitem[Brooks et~al., 2011]{brooks2011handbook}
Brooks, S., Gelman, A., Jones, G., and Meng, X.-L. (2011).
\newblock {\em Handbook of markov chain monte carlo}.
\newblock CRC press.

\bibitem[Chiu et~al., 1996]{chiu1996matrix}
Chiu, T.~Y., Leonard, T., and Tsui, K.-W. (1996).
\newblock The matrix-logarithmic covariance model.
\newblock {\em Journal of the American Statistical Association},
  91(433):198--210.

\bibitem[Eguchi et~al., 2010]{eguchi2010priori}
Eguchi, N., Saito, R., Saeki, T., Nakatsuka, Y., Belikov, D., and Maksyutov, S.
  (2010).
\newblock A priori covariance estimation for co2 and ch4 retrievals.
\newblock {\em Journal of Geophysical Research: Atmospheres}, 115(D10).

\bibitem[Fan et~al., 2008]{fan2008high}
Fan, J., Fan, Y., and Lv, J. (2008).
\newblock High dimensional covariance matrix estimation using a factor model.
\newblock {\em Journal of Econometrics}, 147(1):186--197.

\bibitem[Gelfand and Smith, 1990]{gelfand1990sampling}
Gelfand, A.~E. and Smith, A.~F. (1990).
\newblock Sampling-based approaches to calculating marginal densities.
\newblock {\em Journal of the American statistical association},
  85(410):398--409.

\bibitem[Gelman and Hill, 2006]{gelman2006data}
Gelman, A. and Hill, J. (2006).
\newblock {\em Data analysis using regression and multilevel/hierarchical
  models}.
\newblock Cambridge university press.

\bibitem[Gelman et~al., 2013]{gelman2013bayesian}
Gelman, A., Stern, H.~S., Carlin, J.~B., Dunson, D.~B., Vehtari, A., and Rubin,
  D.~B. (2013).
\newblock {\em Bayesian data analysis}.
\newblock Chapman and Hall/CRC.

\bibitem[Guillot et~al., 2015]{guillot2015statistical}
Guillot, D., Rajaratnam, B., Emile-Geay, J., et~al. (2015).
\newblock Statistical paleoclimate reconstructions via markov random fields.
\newblock {\em The Annals of Applied Statistics}, 9(1):324--352.

\bibitem[Hamimeche and Lewis, 2009]{hamimeche2009properties}
Hamimeche, S. and Lewis, A. (2009).
\newblock Properties and use of cmb power spectrum likelihoods.
\newblock {\em Physical Review D}, 79(8):083012.

\bibitem[Hamsici and Martinez, 2007]{hamsici2007spherical}
Hamsici, O.~C. and Martinez, A.~M. (2007).
\newblock Spherical-homoscedastic distributions: The equivalency of spherical
  and normal distributions in classification.
\newblock {\em Journal of Machine Learning Research}, 8(7).

\bibitem[Hoff, 2009]{hoff2009hierarchical}
Hoff, P.~D. (2009).
\newblock A hierarchical eigenmodel for pooled covariance estimation.
\newblock {\em Journal of the Royal Statistical Society: Series B (Statistical
  Methodology)}, 71(5):971--992.

\bibitem[Hoff and Niu, 2012]{hoff2012covariance}
Hoff, P.~D. and Niu, X. (2012).
\newblock A covariance regression model.
\newblock {\em Statistica Sinica}, pages 729--753.

\bibitem[Kundu and Pourahmadi, 2020]{kundu2020mle}
Kundu, A. and Pourahmadi, M. (2020).
\newblock Mle of jointly constrained mean-covariance of multivariate normal
  distributions.
\newblock {\em arXiv preprint arXiv:2012.11826}.

\bibitem[Lange, 2013]{lange2013elementary}
Lange, K. (2013).
\newblock Elementary optimization.
\newblock In {\em Optimization}, pages 1--17. Springer.

\bibitem[Ledoit and Wolf, 2004a]{ledoit2004honey}
Ledoit, O. and Wolf, M. (2004a).
\newblock Honey, i shrunk the sample covariance matrix.
\newblock {\em The Journal of Portfolio Management}, 30(4):110--119.

\bibitem[Ledoit and Wolf, 2004b]{ledoit2004well}
Ledoit, O. and Wolf, M. (2004b).
\newblock A well-conditioned estimator for large-dimensional covariance
  matrices.
\newblock {\em Journal of multivariate analysis}, 88(2):365--411.

\bibitem[Mardia and Jupp, 2000]{mardia2000directional}
Mardia, K. and Jupp, P. (2000).
\newblock Directional statistics.

\bibitem[Murphy, 2012]{murphy2012machine}
Murphy, K.~P. (2012).
\newblock {\em Machine learning: a probabilistic perspective}.
\newblock MIT press.

\bibitem[Paine et~al., 2018]{paine2018elliptically}
Paine, P., Preston, S.~P., Tsagris, M., and Wood, A.~T. (2018).
\newblock An elliptically symmetric angular gaussian distribution.
\newblock {\em Statistics and Computing}, 28(3):689--697.

\bibitem[Paine et~al., 2020]{paine2020spherical}
Paine, P.~J., Preston, S., Tsagris, M., and Wood, A.~T. (2020).
\newblock Spherical regression models with general covariates and anisotropic
  errors.
\newblock {\em Statistics and Computing}, 30(1):153--165.

\bibitem[Pourahmadi, 1999]{pourahmadi1999joint}
Pourahmadi, M. (1999).
\newblock Joint mean-covariance models with applications to longitudinal data:
  Unconstrained parameterisation.
\newblock {\em Biometrika}, 86(3):677--690.

\bibitem[Preston and Paine, 2017]{preston2017analysis}
Preston, S. and Paine, P. (2017).
\newblock {\em Analysis of spherical data with ESAG}.

\bibitem[Rao, 1973]{rao1973linear}
Rao, C.~R. (1973).
\newblock {\em Linear statistical inference and its applications}.
\newblock Wiley New York, second edition.

\bibitem[Sch{\"a}fer and Strimmer, 2005]{schafer2005shrinkage}
Sch{\"a}fer, J. and Strimmer, K. (2005).
\newblock A shrinkage approach to large-scale covariance matrix estimation and
  implications for functional genomics.
\newblock {\em Statistical applications in genetics and molecular biology},
  4(1).

\bibitem[Schmidt, 1976]{schmidt1976non}
Schmidt, P. (1976).
\newblock The non-uniqueness of the australian mesozoic palaeomagnetic pole
  position.
\newblock {\em Geophysical Journal International}, 47(2):285--300.

\bibitem[Trefethen and Bau~III, 1997]{trefethen1997numerical}
Trefethen, L.~N. and Bau~III, D. (1997).
\newblock {\em Numerical linear algebra}, volume~50.
\newblock Siam.

\bibitem[Yang and Berger, 1994]{yang1994estimation}
Yang, R. and Berger, J.~O. (1994).
\newblock Estimation of a covariance matrix using the reference prior.
\newblock {\em The Annals of Statistics}, pages 1195--1211.

\end{thebibliography}

\newpage
\appendix
\appendixpage

\section{Proofs of Results:}
\label{PrRes}
\begin{enumerate}

\item \label{MAPIWPr} In case of normal-inverse Wishart prior the posterior density with parameter $(\bm{\mu}_n,\kappa_n,\nu_n,\bm{\Lambda}_n)$ is the following:

\begin{align*}
p(\bm{\mu},\bm{\Sigma}\mid\bm{\mu}_n,\kappa_n,\nu_n,\bm{\Lambda}_n) &\propto \kappa_n^{-\frac{p}{2}}\mid\bm{\Sigma}\mid^{-\left(\frac{\nu_n+p}{2}+1\right)} \exp\left[-\frac{1}{2}\left\{\text{Tr}\left(\bm{\Sigma}^{-1}\bm{\Lambda}_n\right)+\kappa_n\left(\bm{\mu}-\bm{\mu}_n\right)^{\top}\bm{\Sigma}^{-1}\left(\bm{\mu}-\bm{\mu}_n\right)\right\}\right]
\end{align*}

After taking logarithm and differentiating we get :

\begin{align*}
\frac{\partial \log p(\bm{\mu},\bm{\Sigma}\mid\bm{\mu}_n,\kappa_n,\nu_n,\bm{\Lambda}_n)}{\partial \bm{\Sigma}^{-1}}&=\left(\frac{\nu_n+p}{2}+1\right)\bm{\Sigma} -\frac{1}{2}\bm{\Lambda}_n -\frac{\kappa_n}{2}\left(\bm{\mu}-\bm{\mu}_n\right)\left(\bm{\mu}-\bm{\mu}_n\right)^{\top} = 0\\
\widehat{\bm{\Sigma}}_{map} &= \frac{\bm{\Lambda}_n+\kappa_n\left(\bm{\mu}-\bm{\mu}_n\right)\left(\bm{\mu}-\bm{\mu}_n\right)^{\top}}{\nu_n + p + 2}
\end{align*}

By applying Theorem 4.2.1 of \cite{bibby1979multivariate} we can say that this is indeed the MAP estimator.

\item \label{NSIWConPr}
 Let $\bm{x}_1,\bm{x}_2,\dots,\bm{x}_n\sim N_p(\bm{\mu},\bm{\Sigma})$ and the prior on $(\bm{\mu},\bm{\Sigma})$ is normal-shrinkage inverse Wishart prior i.e. $\bm{\mu}\mid\bm{\Sigma}\sim N_p(\bm{\mu}_0,\bm{\Sigma}/\kappa_0)$ and $\bm{\Sigma}\sim \pi_{SIW}(\nu_0,b,\bm{\Lambda}_0^{-1})$. Here w are interested with $b=1$ as stated earlier. The posterior distribution is calculated as follows:

\begin{align*}
\pi(\bm{\mu},\bm{\Sigma}\mid \bm{x}_1,\dots,\bm{x}_n,\bm{\mu}_0,\kappa_0,\nu_0,\bm{\Lambda}_0) &\propto \prod_{i=1}^n N_p(\bm{x}_i; \bm{\mu},\bm{\Sigma}) N_p(\bm{\mu}; \bm{\mu}_0,\bm{\Sigma}/\kappa_0) \pi_{SIW}(\bm{\Sigma};\nu_0,1,\bm{\Lambda}_0^{-1}) \\
&\propto  \prod_{i=1}^n N_p(\bm{x}_i; \bm{\mu},\bm{\Sigma}) N_p(\bm{\mu}; \bm{\mu}_0,\bm{\Sigma}/\kappa_0) \frac{\pi_{IW}(\bm{\Sigma};\nu_0,\bm{\Lambda}_0^{-1})}{\prod_{i<j}(\lambda_i-\lambda_j)}
\end{align*}

\noindent This is because shrinkage-inverse-Wishart can be obtained from inverse-Wishart density by dividing it with $\prod_{i<j}(\lambda_i - \lambda_j)$.  We know that the normal-inverse Wishart prior is conjugate with posterior parameters described in (\ref{PostPar}), $\kappa_n = \kappa_0+n$ and $\nu_n=\nu_0+n$ respectively. The numerator is the posterior distribution corresponding to the normal-inverse Wishart prior and gives us the following:

\begin{align*}
\pi(\bm{\mu},\bm{\Sigma}\mid \bm{x}_1,\dots,\bm{x}_n,\bm{\mu}_0,\kappa_0,\nu_0,\bm{\Lambda}_0) &\propto\frac{ \pi_{NIW}(\bm{\mu},\bm{\Sigma}\mid \bm{\mu}_n,\kappa_n,\nu_n,\bm{\Lambda}_n^{-1})}{\prod_{i<j}(\lambda_i-\lambda_j)}\\
&\propto N_p(\bm{\mu}\mid\bm{\Sigma},\bm{\mu}_n,\kappa_n) \frac{\pi_{IW}(\bm{\Sigma}\mid\nu_n,\bm{\Lambda}_n^{-1})}{\prod_{i<j}(\lambda_i-\lambda_j)}\\
&\propto N_p(\bm{\mu}\mid\bm{\Sigma},\bm{\mu}_n,\kappa_n) \pi_{SIW}(\bm{\Sigma}\mid\nu_n,1,\bm{\Lambda}_n^{-1})
\end{align*}

From the above calculation we can see that the posterior is normal-shrinkage inverse Wishart with the same parameters as in normal-inverse Wishart proving our hypothesis.

\item \textbf{Calculation of the Posterior Distribution:} \begin{align*}
p(\bm{\mu},\bm{\lambda} \mid \bm{X})
&\propto \left(\prod_{i=1}^{p-1} \lambda_i\right)^{-\frac{n}{2}} \exp\left[-\frac{1}{2} Tr\left\{\bm{\Lambda}^{-1} B(\bm{X},\bm{\mu})\right\}\right] \left(\prod_{i=1}^{p-1} \lambda_i\right)^{-\frac{1}{2}} \\
& \quad\quad\quad \exp\left[-\frac{1}{2} Tr\left\{\bm{\Lambda}^{-1} (\bm{\mu}-\bm{\mu}_0)(\bm{\mu} - \bm{\mu}_0)^{\top}\right\}\right] \left(\prod_{i=1}^{p-1} \lambda_i\right)^{-a}   \exp\left[-\frac{1}{2} Tr\left\{\bm{\Lambda}^{-1}\bm{H}_0\right\}\right]\\
&\propto \left(\prod_{i=1}^{p-1} \lambda_i\right)^{-\frac{n+2(a+1)}{2}} \exp\left[-\frac{1}{2}Tr\left\{\bm{\Lambda}^{-1}\left(B(\bm{X},\bm{\mu}) + (\bm{\mu}-\bm{\mu}_0)(\bm{\mu} - \bm{\mu}_0)^{\top} + \bm{H}_0 \right)\right\}\right]\\
 &\propto   \left(\prod_{i=1}^{p-1} \lambda_i\right)^{-\frac{n+2(a+1)}{2}}  \exp\left[-\frac{1}{2}Tr\left\{\bm{\Lambda}^{-1} \bm{H}_N\right\}\right]
\end{align*}
 where $B(\bm{X},\bm{\mu}) = \bm{P}(\bm{\mu})^{\top} \bm{A}(\bm{\mu}) \bm{P}(\bm{\mu})$.

\end{enumerate}

\end{document}